\documentclass{article}

\PassOptionsToPackage{numbers,sort&compress}{natbib}
\usepackage[preprint]{neurips_2026}

\usepackage{makecell}
\usepackage[utf8]{inputenc} 
\usepackage[T1]{fontenc}    
\usepackage{hyperref}       
\usepackage{url}            
\usepackage{booktabs}       
\usepackage{amsfonts}       
\usepackage{nicefrac}       
\usepackage{microtype}      
\usepackage{xcolor}         
\usepackage{graphicx} 
\usepackage{booktabs}
\usepackage{multirow}
\usepackage{amsmath}
\usepackage{graphicx} 
\usepackage{booktabs}
\usepackage{pifont}
\usepackage{amssymb}
\usepackage{tabularx}

\newcommand{\bcmark}{\text{\boldmath\ding{51}}}

\title{Multi-site PPG: An In-the-Wild  Physiological Dataset from Emerging Multi-site Wearables}

\author{%
{\bf Jiayi Shao$^{1}$, Jiaying Ye$^{1}$, Shengyao Liu$^{1}$, Zachary Englhardt$^{1}$, Girish Narayanswamy$^{1}$} \\
{\bf Vikram Iyer$^{1}$, Qiuyue (Shirley) Xue$^{2}$} \\
$^{1}$University of Washington \quad
$^{2}$Purdue University \\
{\small
\texttt{\{jyshao34,yjy0509,sliu1229, zacharye, girishvn, vsiyer\}@uw.edu}, \texttt{qiuyue@purdue.edu}
}
}

\begin{document}

\maketitle

\begin{abstract}

Wearables are widely used for mobile health monitoring, and photoplethysmography (PPG) is a key sensing modality for heart rate and related physiological measurements. However, public in-the-wild PPG datasets remain largely wrist-centric or limited to short, controlled studies, constraining research on emerging wearable form factors. We present Multi-site PPG, an in-the-wild physiological dataset collected from four custom-developed unobtrusive wearables: a smart earring, ring, watch, and necklace. Each device records green and infrared reflective PPG, 3-axis acceleration, and temperature with timestamps for cross-device alignment, while a Polar H10 chest strap provides reference electrocardiogram (ECG). Participants wore the devices for one to multiple days during daytime activities while continuing their normal routines. The dataset contains over 350 hours of raw data and 230–290 hours of modeling-ready 8-second windows per wearable. We benchmark heuristic, supervised, and self-supervised heart-rate estimation methods, showing substantial body-site differences: the best methods achieve mean absolute errors (MAEs) of 2.30 bpm on the earring, 5.13 bpm on the ring, 8.37 bpm on the watch, and 8.68 bpm on the necklace. We further analyze motion effects and evaluate multi-site and PPG–accelerometer fusion, demonstrating the dataset’s value for robust physiological sensing across emerging wearable form factors.

\textcolor{blue}{Dataset: } 
\href{https://huggingface.co/datasets/snowballlab/Multisite-PPG}{https://huggingface.co/datasets/snowballlab/Multisite-PPG}

\textcolor{blue}{Code: } 
\href{https://github.com/jiayimaggieshao/wearable-ppg-dataset}{https://github.com/jiayimaggieshao/wearable-ppg-dataset}

\end{abstract}

\section{Introduction}

\begin{figure}[ht]
    \centering
    \includegraphics[width=0.98\linewidth]{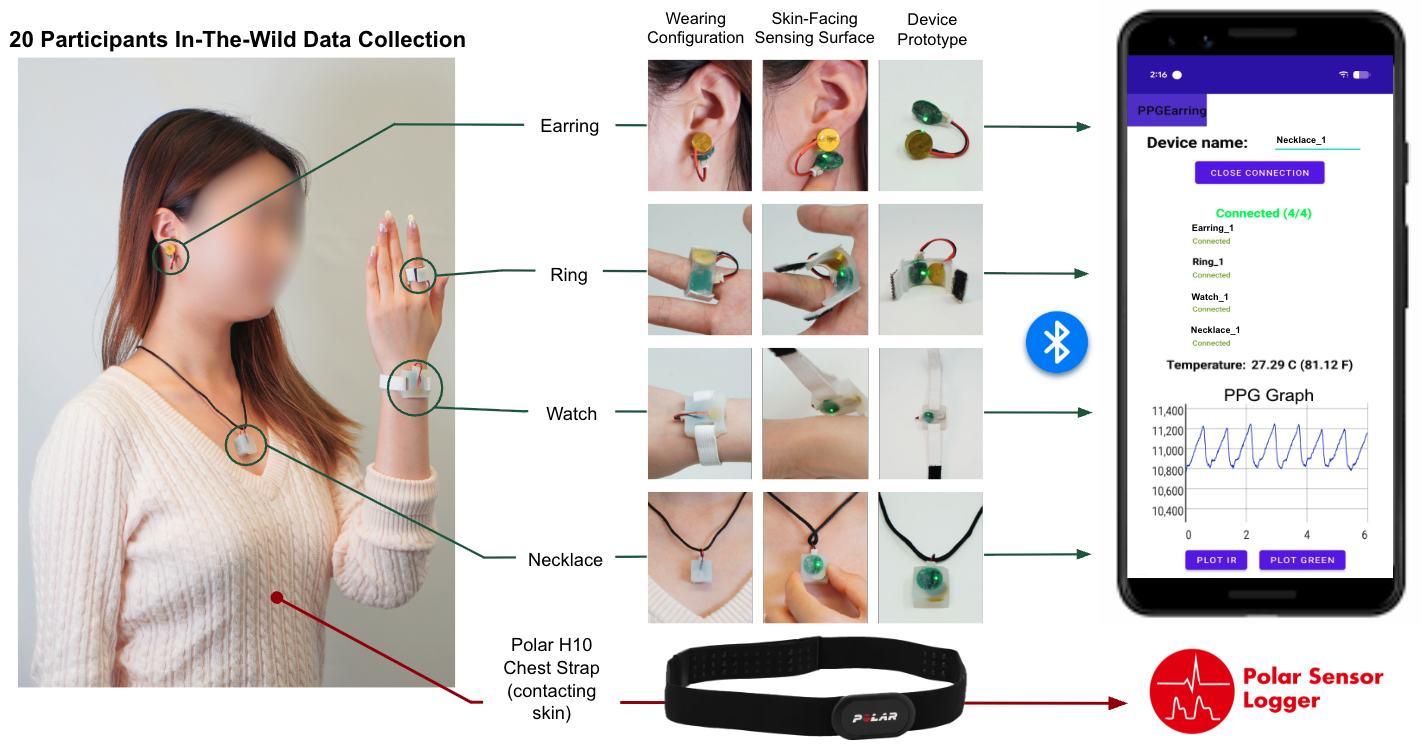}
    \caption{\small{Multi-site data collection setup. Our dataset collects signals from four wearable devices: earring, ring, watch, necklace. Each device shares the same sensing hardware and streams two-channel reflective PPG (green and IR), 3-axis accelerometer, and temperature data to a smartphone app. A Polar H10 chest strap provides reference ECG for ground truth.}}
    \label{fig:study_setup}
    \vskip -1em
\end{figure}

Photoplethysmography (PPG) is widely used in wearable health monitoring for heart rate estimation, respiration sensing, and longitudinal physiological tracking. However, most public wearable PPG datasets are collected in short, controlled settings or focus primarily on smartwatch data. Although convenient, smartwatch sensing on the wrist is challenging: signals are often degraded by motion and lower optical signal quality because major blood vessels lie deeper beneath the skin. As a result, wrist PPG systems often struggle during daily activities, especially for richer physiological signals such as heart rate variability and breathing rate.

Emerging wearable form factors offer an opportunity to move beyond wrist-centric PPG sensing. Smart rings are increasingly used for consumer health monitoring, and the finger provides strong PPG signals due to high blood perfusion~\cite{oura, cao2022accuracy, mehrabadi2020sleep}. The earlobe is also promising because of its rich perfusion, thin tissue, and relative stability; recent smart earring prototypes suggest that it can support high-quality, motion-resistant PPG in daily life~\cite{xue2025ppg}.

Despite this promise, public in-the-wild PPG datasets from emerging wearable form factors remain scarce. Existing smart ring PPG datasets are typically limited to short, controlled lab sessions, such as the 28-hour $\tau$-Ring dataset~\cite{tang2025dataset}. Beyond single-device recordings, multi-site PPG datasets are needed to quantify the benefits of non-wrist sensing, characterize physiological and signal-quality differences across body locations, evaluate multi-site sensor fusion, and develop models that generalize across wearable form factors. However, such datasets remain limited because they are technically challenging to collect: emerging wearables such as rings impose strict constraints on size, weight, battery life, and user comfort. WildPPG~\cite{meier2024wildppg} is a closely-related dataset that includes multiple body sites, but uses smartwatch-style devices placed on locations such as the sternum and ankle, rather than practical everyday form factors such as rings and earrings.

To address this gap, we present \textbf{Multi-site PPG: an in-the-wild physiological dataset from emerging wearables}. As shown in Figure~\ref{fig:study_setup}, the dataset was collected during naturalistic daytime activities using four custom-developed unobtrusive wearables: a smart earring, ring, watch, and necklace. Each device streams PPG, accelerometer, and temperature signals to a smartphone app with timestamps for later alignment. Participants used the app to self-report their activities such as studying, walking, and exercising. A commercial Polar chest strap provides reference electrocardiogram (ECG) signals, from which we derive ground-truth heart rate. Our release contains a raw dataset with over 350 hours per wearable site, and a modeling-ready windowed dataset with 229.66–290.23 hours per site. 



We benchmark heart rate estimation against ECG-derived ground truth using six heuristic algorithms, five supervised models, and two self-supervised models. With the best method, the smart earring achieves an average mean absolute error (MAE) of \textbf{2.30 bpm}, while the ring achieves \textbf{5.13 bpm}, comparable to the 5.18 bpm reported by the 28-hour $\tau$-Ring dataset~\citep{tang2025dataset}. The watch achieves \textbf{8.37 bpm}, comparable to results reported by WildPPG~\cite{meier2024wildppg}, and the necklace achieves \textbf{8.68 bpm}. These results show substantial performance differences across body sites and highlight the potential of emerging form factors, particularly ring and earring, for robust PPG sensing beyond the watch. 

We further evaluate multi-site and multimodal modeling. Two-device fusion provides moderate improvements, suggesting complementary information across body sites, whereas four-device fusion does not yield additional gains. Combining PPG with accelerometer z-axis data consistently improves performance for the ring and watch, but provides limited benefit for the earring and necklace. Overall, these results demonstrate that the dataset supports a broad range of future research on wearable PPG, including multi-site modeling, multimodal fusion, and multi-wavelength sensing.

In summary, this paper makes three main contributions:
\vskip -1em
\begin{itemize}
\item \textbf{A public in-the-wild multi-site PPG dataset.} We release a 230+ hour physiological dataset collected during daily activities from four simultaneous wearable form factors: earring, ring, watch, and necklace. To the best of our knowledge, it is the first large-scale in-the-wild PPG dataset for smart rings and the first public PPG dataset for earrings and necklaces.

\item \textbf{Body-location benchmarking.} We benchmark heuristic, supervised, and self-supervised heart-rate estimation methods against ECG-derived ground truth, showing clear site-dependent performance differences, with MAEs of 2.30 bpm for the earring, 5.13 bpm for the ring, 8.37 bpm for the watch, and 8.68 bpm for the necklace. We further analyze how motion affects estimation accuracy.

\item \textbf{Multi-site and multimodal modeling.} We evaluate cross-site and PPG--accelerometer fusion, demonstrating the dataset's value for studying complementary body-site sensing and robust wearable health modeling.

\end{itemize}



\section{Related Work}

\begin{table*}[h]
\centering
\caption{Comparison of related wearable PPG datasets.}
\label{tab:dataset_comparison}
\small
\setlength{\tabcolsep}{3pt}
\renewcommand{\arraystretch}{1.0}
\begin{tabular}{l l c l l}
\toprule
\textbf{Dataset} &
\textbf{In-the-wild} &
\textbf{Hours} &
\textbf{Wearable form factor} &
\textbf{Multi-site PPG} \\
\midrule

\textbf{Multi-site PPG (ours)} &
\bcmark &
\textbf{230+} &
\textbf{Earring, Ring, Watch, Necklace} &
\bcmark \\

$\tau$-Ring Dataset~\citep{tang2025dataset} &
Controlled &
28.2 &
Ring &
$\times$ \\

WF-PPG~\citep{ho2025wf} &
Controlled &
~10 &
Non-wearable (clamp) &
\checkmark \\

GalaxyPPG~\citep{park2025galaxyppg} &
Semi-natural &
$\sim$15 &
Watch  &
$\times$ \\

WildPPG~\citep{meier2024wildppg} &
\checkmark &
216 &
Watch and non-everyday placements &
\checkmark \\

EarSet~\citep{montanari2023earset} &
Controlled &
17 &
Earbuds &
$\times$ \\

DaLiA~\citep{ppg-dalia_495} &
\checkmark &
36 &
Watch &
$\times$ \\

BAMI~\citep{lee2018motion} &
Controlled &
5.6 &
Watch &
$\times$ \\

WESAD~\citep{schmidt2018introducing} &
Controlled &
$\sim$30 &
Watch &
$\times$ \\

Wrist PPG Exercise~\citep{jarchi2016description} &
Controlled &
$\sim$2.3 &
Watch &
$\times$ \\

\bottomrule
\end{tabular}
\end{table*}

\subsection{PPG-related Physiological Datasets}

A growing body of work has introduced public PPG datasets, but many are collected in controlled settings over short durations, or primarily from smartwatches. As summarized in Table~\ref{tab:dataset_comparison}, prior datasets include lab-controlled wrist-based datasets for stress monitoring and exercise~\cite{schmidt2018introducing,lee2018motion,chung2020deep,jarchi2016description}. Other datasets target specific tasks or non-wearable settings, such as BIDMC for respiratory-rate estimation~\cite{pimentel2016toward}, smartphone fingertip PPG for waveform quality assessment~\cite{nemcova2021brno,PhysioNet-butppg-2.0.0,neshitov2021wavelet}, and camera-based remote PPG~\cite{mcduff2022scamps}. 

Recent datasets have moved toward more naturalistic smartwatch datasets such as GalaxyPPG and PPG-DaLiA~\cite{park2025ppg,ppg-dalia_495}. However, these datasets remain largely single-site and wrist-centric. Only a few datasets explore multi-site or novel-site PPG: WildPPG provides long recordings from smartwatch-style sensors placed on the wrist and several less practical body locations~\cite{meier2024wildppg}, $\tau$-Ring provides 28 hours of controlled ring PPG~\cite{tang2025dataset}, and EarSet studies in-ear PPG under body motion in short lab-controlled settings~\cite{montanari2023earset}. In contrast, our dataset provides simultaneous in-the-wild PPG recordings from an earring, ring, watch, and necklace, enabling direct comparison across emerging wearables.

\subsection{Physiological Estimation from PPG}


PPG is widely used for heart-rate estimation, pulse-rate variability analysis, and respiratory-rate estimation~\cite{vest2018open}. Classical methods estimate heart rate from spectral peaks, pulse peaks, derivative zero-crossings, or systolic upslopes~\cite{bracewell1989fourier,van2019heartpy,bishop_ercole_2018,charlton2022detecting}, while later work improves robustness to motion using spectrum-based methods~\cite{zhang2014troika, zhang2015photoplethysmography}, neural network models~\cite{ismail2022heart, wilkosz2021multi, bieri2023beliefppg, reiss2019deep}, and PPG–IMU fusion~\cite{kasnesis2022multi, bieri2023beliefppg,kasnesis2023feature}. More recently, self-supervised and foundation-model approaches have learned generalizable PPG representations through contrastive learning, masked reconstruction, denoising, temporal prediction, and cross-modal alignment~\cite{saha2025pulse,zhang2024general,yun2024unsupervised,benfenati2025enhanceppg,ding2024siamquality,narayanswamyscaling,ni2025ppg}. However, these methods are often trained or evaluated on single watch datasets, limiting our understanding of how PPG models generalize across body sites and wearable form factors.
Multi-site PPG sensing has been explored for robust heart-rate estimation and pulse-transit-time analysis because PPG quality and morphology vary across body locations and can reflect vascular dynamics~\cite{meier2024robust,meier2024assessing}. Recent fusion methods also show that dynamically weighting sensors by signal quality can improve heart-rate estimation robustness~\cite{meier2024robust}. 

\section{Multi-site PPG Dataset}
\label{dataset}

\subsection{Wearable Device Setup}

Figure~\ref{fig:study_setup} shows the wearable device setup for our data collection. Participants were asked to wear four custom-developed wearable devices: a smart earring, smart ring, smartwatch, and smart necklace. All devices share the same sensing and hardware design and are covered with silicone to isolate the electronics from direct skin contact, but use different attachment mechanisms for wearing at different body locations. The smart earring is worn on a pierced earlobe, with the battery in front and the sensing PCB behind the earlobe, secured by a friction earring back. The smart ring and smartwatch are mounted with adjustable Velcro bands, while the smart necklace is worn as a pendant on a necklace chain. All four devices wirelessly stream sensor data to a companion smartphone app, allowing participants to move freely during daily activity without cables.  
In addition to the four wearable devices, participants wore a commercial \href{https://www.polar.com/us-en/sensors/h10-heart-rate-sensor}{Polar H10 heart rate monitor chest strap} to provide reference ground-truth heart rate. During the study, we used the Polar Sensor Logger app to collect raw ECG at 130 Hz as ground-truth reference, and Polar accelerometer data at 25 Hz.

\begin{figure}[ht]
    \centering
    \includegraphics[width=1.0\linewidth]{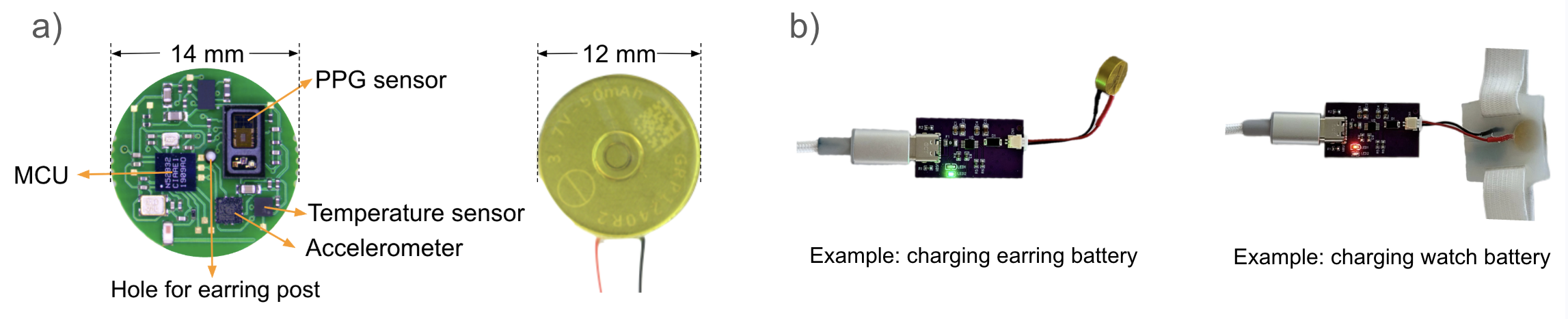}
    \caption{\small{(a) Hardware design of the wearable sensing platform. Each device is built on the same PCB integrating the microcontroller and PPG, accelerometer, and temperature sensor, and is powered by a rechargeable battery. (b) Custom charging adapters shown with the watch and earring form factors.}}
    \label{fig:hardware_charging}
\end{figure}


Figure~\ref{fig:hardware_charging} (a) shows the sensing hardware of our custom wearable devices.  Each device integrates an nRF52832 microcontroller (MCU), MAX30101 PPG sensor, LIS2DW12 accelerometer, and HDC2010 temperature sensor on a printed circuit board (PCB), and is powered by a rechargeable battery GRP1240. To improve electrical safety, comfort, and signal quality, a skin-safe silicone layer covers the PCB surface. This layer compensates for component-height differences, creating a flatter, skin-friendly surface that helps maintain consistent sensor-to-skin contact.  The hardware weighs 1.7 grams before adding the silicone covering, which is within the typical weight range of consumer wearable devices~\cite{apple_watch_series11_specs, oura_ring4}. We have also designed a custom USB charger for the wearable battery, as shown in Figure~\ref{fig:hardware_charging} (b), so the participants can recharge the devices during data collection. 



\subsection{Data Collection Protocol}
\label{protocol}

We conducted a multi-day, in-the-wild data collection in which participants wore multiple wearable devices during their normal daily routines in urban environments. Participants first met the research team in a university laboratory for onboarding, where researchers introduced the data collection protocol and the five devices: smart earring, ring, watch, necklace, and Polar chest strap. After providing consent, participants were shown how to wear each device, use the PPG App and Polar Sensor Logger App on the provided Android smartphone, and recharge the custom wearable devices.

Participants were asked to wear the devices for multiple days during daytime activities of their choice and to continue their normal routines in any setting. They were asked to self-report their activities using the PPG App. For safety and device protection, participants were instructed to remove the devices during sleep or when they might come into contact with water, such as washing hands or bathing. 
Participants were informed of potential risks, including mild discomfort, rare skin irritation, low electrical risk, and privacy risk. The study was approved by the university IRB, and participants were compensated at \$40 per day of device wearing plus \$1 per self-reported activity log.

The dataset contains data from 20 participants, with ages ranging from 20 to 51 years, an average age of $27.9 \pm 8.3$ years, and self-reported Fitzpatrick skin types ranging from I to V. Our dataset has a high representation of females (19 females and 1 male), partly because of the requirement that participants have pierced earlobes. In total, each participant kept the devices for 2 to 6 days, with total wear durations ranging from 6 to 54 hours depending on availability.

\subsection{Dataset Structure}

Our release includes both a raw dataset and an 8-second windowed PPG dataset, which is preprocessed from the raw data and ready for model training.

\begin{table}[t]
\centering
\caption{The synchronized time series captured in our Multi-site PPG Dataset.}
\label{tab:synchronized_time_series}
\setlength{\tabcolsep}{2pt}
\renewcommand{\arraystretch}{0.9}
\small
\begin{tabular}{lllcc}
\toprule
\textbf{Device group} & \textbf{Location} & \textbf{Time Series} & \textbf{Description} & \textbf{Sampling rate} \\
\midrule

\multirow{6}{*}{4 Wearable devices}
& \multirow{6}{*}{\shortstack[l]{Earlobe, Neck\\Finger, Wrist}}
& \texttt{ppg\_g} & Green PPG & 100 Hz \\
&
& \texttt{ppg\_ir} & Infrared PPG & 100 Hz \\
&
& \texttt{temp} & Temperature & 1 Hz \\
&
& \texttt{accel\_x} & Accelerometer X-axis & 100 Hz \\
&
& \texttt{accel\_y} & Accelerometer Y-axis & 100 Hz \\
&
& \texttt{accel\_z} & Accelerometer Z-axis & 100 Hz \\

\midrule
\multirow{4}{*}{\shortstack[l]{Reference device\\(Polar H10 chest strap)}}
& \multirow{4}{*}{Chest}
& \texttt{ecg} & Chest-strap ECG & 130 Hz \\
&
& \texttt{accel\_x} & Accelerometer X-axis & 25 Hz \\
&
& \texttt{accel\_y} & Accelerometer Y-axis & 25 Hz \\
&
& \texttt{accel\_z} & Accelerometer Z-axis & 25 Hz \\

\bottomrule
\end{tabular}
\end{table}

\subsubsection{Raw Dataset} 
The raw dataset contains the original timestamped recordings from the four wearable devices (earring, ring, watch, and necklace) and the Polar H10 chest strap. In total, it includes 355.29 hours of earring data, 360.68 hours of ring data, 367.38 hours of watch data, and 359.80 hours of necklace data from 20 participants, all with aligned Polar reference data. As shown in Table~\ref{tab:synchronized_time_series}, each wearable records green and IR PPG, 3-axis acceleration at 100 Hz, and temperature at 1 Hz, which is interpolated to 100 Hz for timestamp alignment. The Polar chest strap provides raw ECG at 130 Hz and accelerometer data at 25 Hz. Because participants could freely remove and re-wear devices, and because Bluetooth transmission loss occasionally occurred, the raw recordings contain discontinuities. We preserve these non-continuous segments with their original timestamps, making the raw dataset useful for heart-rate estimation as well as missing-data interpolation, imputation, and signal reconstruction.


\subsubsection{Preprocessed Windowed Dataset}
\label{windowed_dataset}
The windowed dataset is derived from the raw dataset using an 8-second sliding window segmentation. Unlike the raw dataset, the windowed dataset contains only continuous segments without sample gaps caused by Bluetooth packet loss or device disconnection. Data from each wearable device (earring, ring, watch, and necklace) are aligned with the Polar ECG ground truth data using the timestamp. We then segment the synchronized data into 8-second windows with a 1-second stride. Each window includes 100 Hz IR PPG, green PPG, and 3-axis accelerometer signals from the wearable device, the synchronized Polar ECG segment, and a ground-truth heart rate label. 

\begin{figure}[ht]
    \centering
    \includegraphics[width=0.9\linewidth]{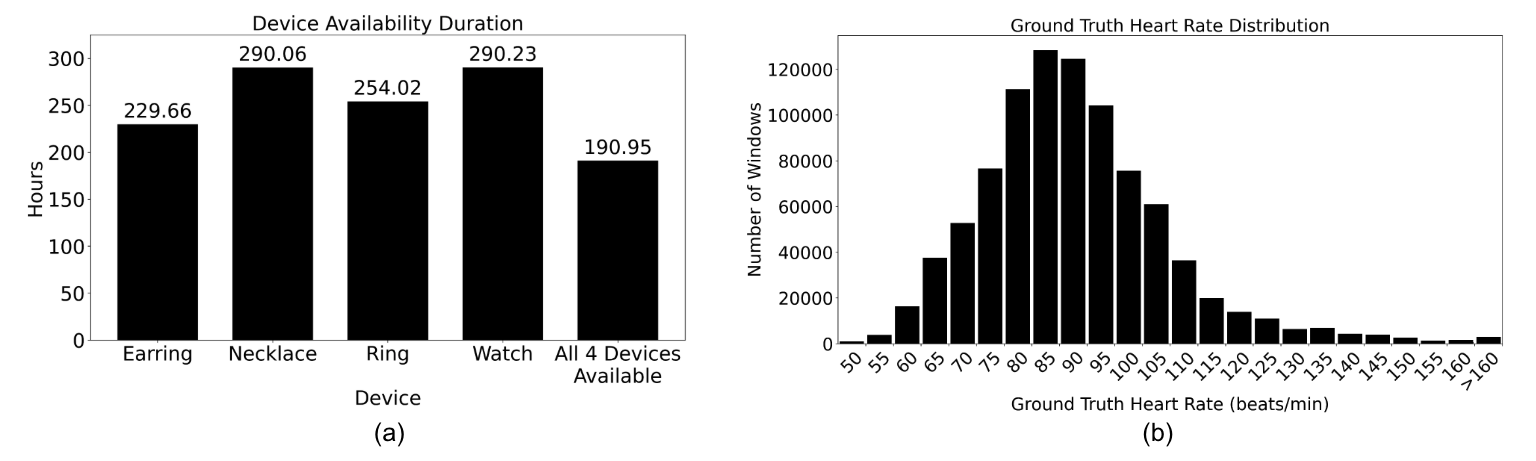}
    \caption{\small{(a) Duration of the dataset collected from each wearable device and the subset in which all four wearable devices are simultaneously available. (b) Distribution of ground-truth heart rate across the dataset.}}
    \label{fig:dataset_distribution}
    \vskip -1em
\end{figure}

As shown in Figure~\ref{fig:dataset_distribution} (a), our 8-second continuous windowed dataset contains 229.66 hours of earring data, 254.02 hours of ring data, 290.23 hours of watch data, and 290.06 hours of necklace data, all with aligned Polar ECG for reference. Among these recordings, 190.95 hours contain simultaneous data from all four wearable devices. This corresponds to 692,577 windows for earring, 774,046 windows for ring, 904,541 windows for watch, 904,550 windows for necklace.

In addition to the dataset, we also provide code for generating additional windowed datasets from the raw dataset with customizable window lengths for different applications. 


\subsubsection{Ground Truth Heart Rate}
Ground-truth heart rate label for each segment in the windowed dataset is computed from the Polar ECG segment using the standard Pan–Tompkins algorithm~\citep{pan1985real}. Figure~\ref{fig:dataset_distribution} (b) shows the ground-truth heart rate distribution in our dataset. To ensure heart rate label quality, the ECG windows are screened to reject segments with invalid or poor-quality ECG, usually caused by loose chest-strap contact. Specifically, we discard windows that contain missing ECG values, are identified as poor quality by the NeuroKit ECG quality function, produce heart rates outside the normal range of 40–180 bpm, or show highly irregular beat intervals within the window, as identified by an excessively large standard deviation of beat-to-beat intervals. During construction of the device-specific 8-second windowed datasets, Polar ECG quality filtering removed 137,146--175,433 windows for each device's dataset, corresponding to 15.59\%--16.53\% of the candidate windows (most of these rejected ECG windows are from 2 participants). The windowed dataset lengths reported in Section~\ref{windowed_dataset} are after removing unqualified ECG windows.




\subsection{Device Synchronization}
Our dataset includes four wearable devices and a Polar chest strap, all streaming data to the same smartphone. We align these data streams using timestamps based on the phone’s clock. Each wearable device keeps its own internal clock and sends this timing information with every Bluetooth packet. When the phone receives the first Bluetooth packet from a device, it maps the device’s internal clock (in microsecond format) to the phone’s clock. This reduces timing errors from Bluetooth transmission and phone processing delays, enabling alignment for the 8-second windows used in our benchmarks. This timestamp-based approach is sufficient for window-level HR estimation, but not for pulse-arrival-time or pulse-transit-time analysis without additional calibration.

\section{Baseline Evaluation}
\label{evaluation}


\begin{table*}[h]
\centering
\caption{Baseline heart-rate estimation results on the four-device Multi-site PPG Dataset, with WildPPG~\citep{meier2024wildppg} included as a contextual reference. WildPPG results, except NeuroKit, are taken from the original publication. Lower MAE/RMSE and higher Pearson correlation ($\rho$) indicate better performance. Values for our dataset are averaged over 20 LOSO folds; standard deviations are reported in Appendix Table~\ref{tab:main_results_supplementary}.}
\label{tab:main_results}
\renewcommand{\arraystretch}{1.15}
\resizebox{\textwidth}{!}{%
\begin{tabular}{lccc|ccc|ccc|ccc|ccc}
\toprule
\multirow{3}{*}{Method}
& \multicolumn{12}{c|}{\textbf{Multi-site PPG Dataset}}
& \multicolumn{3}{c}{WildPPG Dataset} \\
\cmidrule(lr){2-13} \cmidrule(lr){14-16}
& \multicolumn{3}{c|}{\textbf{Earring}}
& \multicolumn{3}{c|}{\textbf{Ring}}
& \multicolumn{3}{c|}{\textbf{Watch}}
& \multicolumn{3}{c|}{\textbf{Necklace}}
& \multirow{2}{*}{MAE$\downarrow$}
& \multirow{2}{*}{RMSE$\downarrow$}
& \multirow{2}{*}{$\rho\uparrow$} \\
\cmidrule(lr){2-4} \cmidrule(lr){5-7} \cmidrule(lr){8-10} \cmidrule(lr){11-13}
& MAE$\downarrow$ & RMSE$\downarrow$ & $\rho\uparrow$
& MAE$\downarrow$ & RMSE$\downarrow$ & $\rho\uparrow$
& MAE$\downarrow$ & RMSE$\downarrow$ & $\rho\uparrow$
& MAE$\downarrow$ & RMSE$\downarrow$ & $\rho\uparrow$
& & & \\
\midrule

\multicolumn{16}{l}{\textit{\textbf{Heuristic}}} \\

NeuroKit
& $\mathbf{2.30}$ & $\mathbf{5.28}$ & $\mathbf{0.933}$
& $7.54$ & $13.64$ & $0.471$
& $12.43$ & $18.84$ & $0.221$
& $14.58$ & $20.72$ & $0.178$
& $9.81$ & $15.47$ & $0.358$ \\

HeartPy
& 2.60 & 6.43 & 0.903
& 9.44 & 16.91 & 0.459
& 15.68 & 23.74 & 0.222
& 20.05 & 28.67 & 0.145
& 18.49 & 30.15 & 0.147 \\

qppgfast
& 3.36 & 7.67 & 0.871
& 11.94 & 19.74 & 0.255
& 15.09 & 22.02 & 0.194
& 17.84 & 25.16 & 0.121
& 19.32 & 29.94 & 0.213 \\

MSPTD
& 2.47 & 6.13 & 0.906
& 10.09 & 17.13 & 0.331
& 16.54 & 23.58 & 0.125
& 19.72 & 26.33 & 0.109
& 13.30 & 21.19 & 0.214 \\

PWD
& 2.71 & 6.71 & 0.886
& 11.68 & 19.81 & 0.194
& 17.59 & 25.16 & 0.069
& 19.13 & 25.62 & 0.095
& 11.72 & 19.57 & 0.252 \\

FFT
& 2.67 & 6.98 & 0.875
& 11.93 & 20.55 & 0.238
& 19.19 & 27.00 & 0.093
& 22.46 & 30.01 & 0.098
& 17.62 & 28.68 & 0.121 \\

\midrule

\multicolumn{16}{l}{\textit{\textbf{Supervised}}} \\

1D ResNet
& $2.44$ & $5.35$ & $0.879$
& $5.47$ & $\mathbf{10.10}$ & $\mathbf{0.679}$
& $8.74$ & $13.94$ & $0.439$
& $9.01$ & $13.26$ & $\mathbf{0.524}$
& $\mathbf{8.62}$ & $15.0$ & $\mathbf{0.424}$ \\

DCL
& $2.34$ & $5.63$ & $0.867$
& $\mathbf{5.13}$ & $10.13$ & $0.676$
& $8.50$ & $13.73$ & $0.425$
& $\mathbf{8.68}$ & $\mathbf{13.07}$ & $0.521$
& $8.64$ & $\mathbf{14.73}$ & $0.414$ \\

FCN
& $2.66$ & $5.45$ & $0.882$
& $6.12$ & $10.29$ & $0.650$
& $9.39$ & $13.94$ & $0.423$
& $9.35$ & $13.09$ & $0.497$
& $9.79$ & $15.62$ & $0.357$ \\

CNN-LSTM
& $2.43$ & $5.96$ & $0.858$
& $5.33$ & $10.37$ & $0.661$
& $\mathbf{8.37}$ & $\mathbf{13.55}$ & $\mathbf{0.458}$
& $8.74$ & $13.21$ & $0.497$
& $9.28$ & $15.14$ & $0.370$ \\

Transformer
& $2.79$ & $6.47$ & $0.841$
& $6.71$ & $11.90$ & $0.539$
& $9.74$ & $14.74$ & $0.332$
& $10.83$ & $14.49$ & $0.349$
& $10.06$ & $16.23$ & $0.321$ \\

\midrule
\multicolumn{16}{l}{\textit{\textbf{Self-supervised}}} \\

SimCLR
& $3.76$ & $7.32$ & $0.823$
& $9.10$ & $13.70$ & $0.332$
& $10.72$ & $14.49$ & $0.203$
& $11.25$ & $14.88$ & $0.236$
& $15.75$ & $19.81$ & $0.081$ \\

BYOL
& $3.46$ & $7.26$ & $0.822$
& $7.83$ & $12.44$ & $0.475$
& $11.63$ & $15.31$ & $0.234$
& $11.13$ & $14.53$ & $0.317$
& $14.11$ & $18.42$ & $0.125$ \\

\bottomrule
\end{tabular}%
}
\end{table*}

\subsection{Benchmark Methods}
We benchmark heart-rate estimation on our windowed dataset using three classes of methods: six heuristic algorithms, five supervised neural models, and two self-supervised models. All methods estimate heart rate from an 8-second windowed dataset using PPG green light and are evaluated against aligned Polar ECG-derived ground truth. We report mean absolute error (MAE), root mean square error (RMSE), and Pearson correlation coefficient ($\rho$), with results averaged over 20 leave-one-subject-out (LOSO) folds. Table~\ref{tab:main_results} summarizes the results and includes WildPPG~\citep{meier2024wildppg} as a reference dataset for comparison. For fair comparison and reproducibility, the supervised backbone implementations are adapted from the public WildPPG codebase~\cite{meier2024wildppg}, with modifications for our dataset format and evaluation protocol. Additional baseline details are provided in Appendix~\ref{baseline_details}.



\subsection{Results Across Wearables}

\begin{figure}[ht]
    \centering
    \includegraphics[width=1\linewidth]{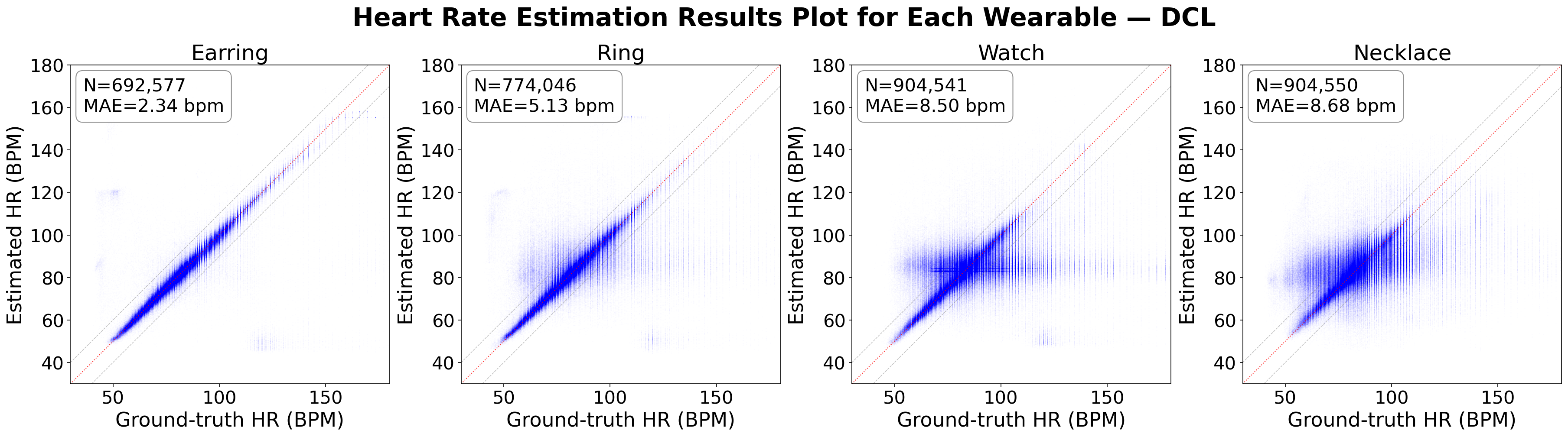}
    \caption{\small{Heart rate estimation results across four wearable devices, using the DCL model.}}
    \label{fig:baseline_results}
    \vskip -1em
\end{figure}

As shown in Table~\ref{tab:main_results}, performance shows a consistent location-dependent trend across heuristic, supervised, and self-supervised methods: \textbf{earring > ring > watch > necklace}. The earring achieves the best MAE of \textbf{2.30} bpm, followed by the ring at \textbf{5.13} bpm, which is comparable to the 5.18 bpm reported by the 28-hour $\tau$-Ring Dataset~\citep{tang2025dataset}. The watch reaches \textbf{8.37} bpm, consistent with results from the watch-worn WildPPG dataset~\citep{meier2024wildppg}, while the necklace performs slightly worse at \textbf{8.68} bpm.

Figure~\ref{fig:baseline_results} visualizes the DCL predictions across the four wearable sites. Each point corresponds to one 8-second window, with ECG-derived heart rate on the x-axis and estimated heart rate on the y-axis. The earring shows the tightest clustering around the identity line, indicating the most accurate and consistent estimates. The ring remains well aligned but exhibits greater dispersion, while the watch and necklace show wider scatter and stronger regression toward a narrower heart-rate range.

These differences likely reflect both physiological and mechanical factors. The earlobe and finger provide stronger perfusion and clearer optical pulse signals, while the earring further benefits from lower motion at the sensing site. In contrast, the wrist is more affected by motion and has weaker optical signal quality, and the necklace suffers from unstable skin contact due to its pendant form factor. Together, these results show that the dataset supports not only HR benchmarking, but also analysis of location-dependent PPG reliability in daily life.

\subsection{Heart Rate Error Versus Motion}
\label{motion_effect}

\begin{figure}[ht]
    \centering
    \includegraphics[width=1\linewidth]{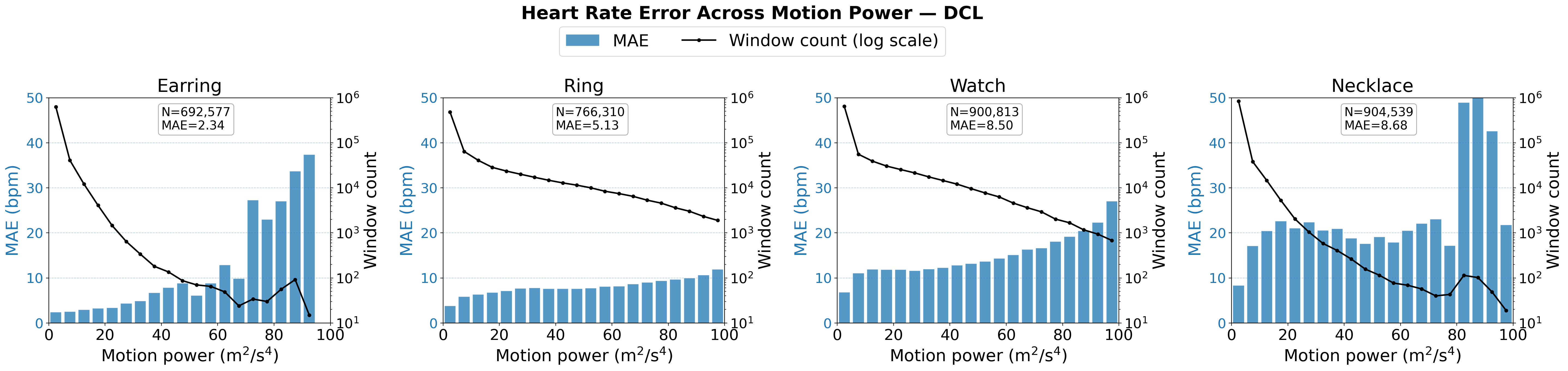}
    \caption{\small{Distribution of windows across accelerometer-derived motion-power bins and corresponding HR estimation MAE for the four device locations using the DCL model.}}
    \label{fig:Error_vs_motion}
    \vskip -1em
\end{figure}

We further examine how motion affects PPG accuracy at each wearable site. Figure~\ref{fig:Error_vs_motion} shows window distributions and heart-rate estimation MAE across motion-power bins. Motion power is computed by removing the mean three-axis acceleration within each 8-second window and averaging the squared magnitude of the remaining dynamic acceleration. The earring and necklace experience lower motion exposure because they are on the head or torso, with over 91\% of windows in the 0--5 motion-power range, while only 62\% of ring windows and 71\% of watch windows fall in this motion range, reflecting more frequent hand and wrist movement.

MAE generally increases with motion power across all devices, confirming that motion artifacts degrade PPG-based HR estimation. In the 0–5 motion range, MAE is lowest for the earring at 2.4 bpm, followed by the ring at 3.8 bpm, the watch at 6.8 bpm, and the necklace at 8.4 bpm. The ring remains more accurate than the watch across motion levels, likely due to stronger finger PPG signals and better optical coupling. High-motion bins above 60 should be interpreted cautiously for the earring and necklace because they contain fewer than 100 windows (0.01\%), whereas the corresponding ring and watch bins contain about 35,000 windows (4.5\%). Overall, these results show site-dependent motion sensitivity and motivate multi-site sensing beyond the wrist. We also provide an initial analysis across self-reported Fitzpatrick skin types in Appendix~\ref{results_across_skintone}.


\section{Multi-site and Multimodal}
\label{multisite_modeling}

\subsection{Multi-site Modeling}

\begin{table*}[h]
\centering
\caption{Multi-site PPG fusion results using 1D ResNet. In the table, E, R, W, and N denote earring, ring, watch, and necklace. Values are the mean $\pm$ standard deviation across 20 LOSO folds.}

\label{tab:multisite_fusion}
\scriptsize
\setlength{\tabcolsep}{4pt}
\resizebox{\textwidth}{!}{
\begin{tabular}{lcccccccc}
\toprule
& \multicolumn{4}{c}{\textbf{Single device}} 
& \multicolumn{3}{c}{\textbf{Two-device fusion}} 
& \textbf{4-device} \\
\cmidrule(lr){2-5} \cmidrule(lr){6-8} \cmidrule(lr){9-9}
\textbf{Metric} &
\textbf{Earring} & \textbf{Ring} & \textbf{Watch} & \textbf{Necklace} &
\textbf{E+R} & \textbf{R+W} & \textbf{W+N} &
\textbf{All} \\
\midrule
MAE$\downarrow$
& $1.84\pm0.71$ & $4.55 \pm 2.04$ & $7.03 \pm 2.43$ & $7.73 \pm 2.72$
& $\mathbf{1.82 \pm 0.65}$ & $\mathbf{4.49 \pm 1.97}$ & $\mathbf{6.90 \pm 2.32}$
& $1.89 \pm 0.69$ \\
RMSE$\downarrow$
& $3.19 \pm 1.61$ & $7.85 \pm 3.02$ & $10.59 \pm 3.25$ & $11.17 \pm 3.33$
& $\mathbf{3.18 \pm 1.57}$ & $\mathbf{7.82 \pm 2.89}$ & $\mathbf{10.36 \pm 3.19}$
& $3.23 \pm 1.62$ \\
$\rho\uparrow$
& $0.94 \pm 0.09$ & $0.74 \pm 0.14$ & $0.50 \pm 0.19$ & $0.48 \pm 0.13$
& $\mathbf{0.95 \pm 0.07}$ & $\mathbf{0.73 \pm 0.15}$ & $\mathbf{0.53 \pm 0.18}$
& $0.95 \pm 0.08$ \\
\bottomrule
\end{tabular}
}
\end{table*}

Since we observed substantial performance differences across wearable sites, we conduct an initial analysis of whether combining PPG signals across body locations can benefit heart-rate estimation. Table~\ref{tab:multisite_fusion} reports 1D ResNet results for single-device, two-device, and four-device configurations. These experiments are intended as a demonstration of the dataset’s potential for multi-site modeling rather than an exhaustive study of fusion architectures. The results are benchmarked on the subset where all four wearables are simultaneously available (190 hours), so they are not directly comparable to the full device-specific results in Table~\ref{tab:main_results}. Overall, two-device fusion consistently improves MAE over the corresponding single-device baselines. For example, combining the earring and ring further reduces MAE to 1.82 bpm, outperforming either earring-only or ring-only models, while ring--watch fusion improves over both the ring-only and watch-only settings. Similarly, watch--necklace fusion outperforms both individual locations. These results suggest that multi-site PPG provides complementary information, even when one of the locations is individually weaker.

Four-device fusion, however, does not yield a clear additional gain in this initial setup. Although its performance is close to the best single- and two-device settings, its MAE remains slightly higher than the best earring-only and earring–ring configurations. One possible explanation is that weaker or noisier sites, such as the watch and necklace, may introduce additional variability when combined with stronger sites. Rather than drawing a definitive conclusion about optimal sensor combinations, these results illustrate that the dataset can support future research on site selection, adaptive fusion, and models that account for differing signal quality across wearable locations.


\subsection{PPG-Motion Multimodal Fusion}
\label{ppg_motion_fusion}

\begin{table*}[h]
\centering
\caption{PPG-accelerometer fusion results across wearable devices. Each backbone is evaluated under PPG (green) versus PPG (green) + accelerometer (z-axis) modalities. Numbers are mean$\pm$std across 20 LOSO folds.}
\label{tab:modality_fusion}
\renewcommand{\arraystretch}{1.15}
\resizebox{\textwidth}{!}{%
\begin{tabular}{ll|cc|cc|cc|cc}
\toprule
\multirow{2}{*}{Backbone} & \multirow{2}{*}{Modality}
& \multicolumn{2}{c|}{\textbf{Earring}}
& \multicolumn{2}{c|}{\textbf{Ring}}
& \multicolumn{2}{c|}{\textbf{Watch}}
& \multicolumn{2}{c}{\textbf{Necklace}} \\
\cmidrule(lr){3-4} \cmidrule(lr){5-6} \cmidrule(lr){7-8} \cmidrule(lr){9-10}
& & MAE$\downarrow$ & $\rho\uparrow$
& MAE$\downarrow$ & $\rho\uparrow$
& MAE$\downarrow$ & $\rho\uparrow$
& MAE$\downarrow$ & $\rho\uparrow$ \\
\midrule
\multirow{2}{*}{DCL} & PPG only & $1.65\!\pm\!0.67$ & $0.95\!\pm\!0.07$ & $4.23\!\pm\!2.17$ & $0.74\!\pm\!0.15$ & $6.90\!\pm\!2.37$ & $0.50\!\pm\!0.19$ & \textbf{$7.61\!\pm\!2.86$} & $0.48\!\pm\!0.16$ \\
 & PPG + Accel\_z & $\mathbf{1.61\!\pm\!0.63}$ & $\mathbf{0.95\!\pm\!0.07}$ & $\mathbf{4.18\!\pm\!1.73}$ & $\mathbf{0.75\!\pm\!0.14}$ & $\mathbf{6.58\!\pm\!2.09}$ & $\mathbf{0.57\!\pm\!0.17}$ & $7.91\!\pm\!3.34$ & \textbf{$0.52\!\pm\!0.15$} \\
\midrule
\multirow{2}{*}{CNN-LSTM} & PPG only & \textbf{$1.67\!\pm\!0.67$} & $0.95\!\pm\!0.07$ & \textbf{$4.42\!\pm\!2.17$} & $0.72\!\pm\!0.15$ & $6.96\!\pm\!2.36$ & $0.49\!\pm\!0.19$ & $\mathbf{7.59\!\pm\!2.85}$ & $\mathbf{0.47\!\pm\!0.13}$ \\
 & PPG + Accel\_z & $1.68\!\pm\!0.66$ & \textbf{$0.95\!\pm\!0.07$} & $4.42\!\pm\!2.07$ & \textbf{$0.73\!\pm\!0.14$} & \textbf{$6.74\!\pm\!2.23$} & \textbf{$0.55\!\pm\!0.18$} & $7.63\!\pm\!3.14$ & \textbf{$0.51\!\pm\!0.15$} \\
\midrule
\multirow{2}{*}{ResNet1D} & PPG only & \textbf{$1.84\!\pm\!0.71$} & \textbf{$0.94\!\pm\!0.08$} & $4.55\!\pm\!2.04$ & $0.74\!\pm\!0.14$ & $7.03\!\pm\!2.43$ & $0.50\!\pm\!0.19$ & \textbf{$7.73\!\pm\!2.72$} & $0.48\!\pm\!0.13$ \\
 & PPG + Accel\_z & $1.93\!\pm\!0.74$ & $0.94\!\pm\!0.09$ & \textbf{$4.42\!\pm\!1.71$} & \textbf{$0.75\!\pm\!0.13$} & \textbf{$6.74\!\pm\!2.27$} & \textbf{$0.57\!\pm\!0.18$} & $7.79\!\pm\!3.22$ & \textbf{$0.51\!\pm\!0.17$} \\
\bottomrule
\end{tabular}%
}
\end{table*}

We also conduct an initial analysis of whether accelerometer data can provide useful context for PPG-based heart-rate estimation. Table~\ref{tab:modality_fusion} compares models trained with green PPG alone to those trained with green PPG plus accelerometer $z$-axis data across four wearable sites, using the 190-hour subset where all four wearables are available. These experiments are intended to demonstrate the dataset’s potential for multimodal modeling rather than provide a comprehensive study of sensor-fusion strategies. Across the three benchmarked backbones, adding accelerometer data most consistently benefits the watch, with a best improvement of 0.3 bpm, and provides small gains for the ring, with a best improvement of 0.1 bpm. The effect is limited for the earring and mixed for the necklace. Additional explorations of multi-site PPG–motion fusion and green–IR PPG fusion are provided in Appendix~\ref{multisite_ppg_motion_fusion} and ~\ref{multi_wave_ppg_fusion}.  These results suggest that the accelerometer data may provide useful motion context for more motion-sensitive sites such as the wrist and finger, while also motivating future work on more adaptive multimodal fusion methods that account for site-specific signal quality and device motion.


\section{Discussion}

\subsection{Social Impact}
\label{social_impact}
This dataset may support positive societal outcomes by enabling research on robust, unobtrusive wearable physiological sensing in daily life, including heart-rate estimation across body locations and form factors. Although the main paper focuses on green-light PPG, the dataset also includes IR PPG, accelerometer, temperature, and ECG-derived beat-to-beat labels, supporting future work on multimodal, multi-site modeling, HRV, and breathing exploration.

At the same time, wearable physiological data raise privacy and misuse concerns, as PPG, motion, and activity-context signals may reveal sensitive information about participants’ physiology or behavior. The dataset should not be used for participant identification or clinical diagnosis without appropriate validation. To mitigate these risks, we anonymize the released data, document intended and out-of-scope uses, and release the dataset under a research-oriented license.

\subsection{Limitations}
\label{limitation}
Our dataset has several limitations. It includes a high proportion of female participants, partly due to the pierced-earlobe requirement for the smart earring. Participants self-reported Fitzpatrick skin types I--V, with no participants reporting type VI. The windowed dataset contains ground-truth heart-rate labels derived from Polar ECG signals using the Pan-Tompkins algorithm~\cite{pan1985real}; these labels may not be perfect and may introduce errors. These limitations should be considered when interpreting benchmark results and using the dataset for future model evaluation.

\section{Conclusion}

We present Multi-site PPG, a large-scale in-the-wild physiological dataset collected from four practical wearable form factors: earring, ring, watch, and necklace. The dataset includes over 350 hours of raw data per site and 229.66--290.23 hours of processed modeling-ready windowed data per site, including 190.95 hours with all four devices simultaneously available. We benchmark heuristic, supervised, and self-supervised heart-rate estimation methods against Polar ECG-derived ground truth. Results show clear body-site differences, with average MAEs of \textbf{2.30 bpm} for the earring, \textbf{5.13 bpm} for the ring, \textbf{8.37 bpm} for the watch, and \textbf{8.68 bpm} for the necklace. We further analyze motion effects and evaluate multi-site and PPG--accelerometer fusion, demonstrating the dataset’s value for studying robust, complementary wearable sensing. To the best of our knowledge, this is the first large-scale in-the-wild PPG dataset for smart rings, and the first public PPG dataset for smart earrings and necklaces.

\bibliographystyle{plainnat}
\bibliography{reference}

\newpage
\appendix
\section*{Appendix}


\section{Dataset Accessibility}
We host the dataset on Hugging Face in .npz format and provide Python loaders, preprocessing scripts, benchmark code, and metadata documentation. The released assets include both raw timestamped streams and modeling-ready 8-second windows. We have also included a sample dataset (\href{https://huggingface.co/datasets/anonymous-ppg-dataset/multisite-ppg-submission/tree/main/sample_data}{link}) since our whole dataset exceeds 4 GB. The dataset is released under CC BY-NC 4.0 for non-commercial research and educational use, and the accompanying code is released under the GNU General Public License v3.0 (GPL-3.0) License. We will maintain the repository after publication by preserving versioned releases, documenting known issues, and providing scripts to regenerate windowed datasets from the raw recordings.

\section{Experimental Settings}
\label{baseline_details}
This section provides more details on the implementations of our data preparation pipeline, heuristic experiments, supervised and self-supervised experiments. In total, our experiments consume around 21 hours on a MacBook Pro CPU for preparing the windowed dataset and computing all heuristic baselines, and a total of around 320 hours on GPU clusters (NVIDIA H100) for running supervised and self-supervised baselines, multi-site and multimodal experiments.



\subsection{Preparing windowed dataset}

We construct the 8-second windowed dataset from raw wearable recordings and Polar chest strap ECG recordings using a timestamp-aligned pipeline implemented in Python. The raw wearable data consist of timestamped PPG signals, including green and infrared channels, together with synchronized accelerometer and temperature measurements from each sensing location. Before windowing, these wearable streams are merged into per-participant .npz files with sample timestamps represented in epoch milliseconds. The Polar ECG recordings are likewise merged into .npz files on the same epoch millisecond time axis. Window construction therefore operates on wearable and ECG streams that are already aligned in wall clock time. For each participant and sensing location, the pipeline identifies time regions in which PPG and ECG overlap, segments these regions into fixed-length windows, computes ground truth heart rate from the aligned ECG, and retains only windows that pass the ECG-based quality checks. The full pipeline is implemented in \texttt{run\_pipeline.py}, which calls \texttt{run\_windowing\_aligned.py}, \texttt{run\_pan\_tompkins.py}, and \texttt{merge\_alignment\_windows\_dual\_channel.py}.

\textbf{Windowing and alignment.} Windowing is performed separately for each participant and wearable location on the shared epoch time axis. The pipeline first identifies continuous overlap regions between the wearable PPG stream and the ECG reference, and then extracts fixed-length windows of 8s with a stride of 1s. Each PPG window contains 800 samples at 100Hz. For every PPG window, the corresponding ECG segment is obtained by slicing the ECG signal using the same start and end timestamps, rather than by sample index matching. The resulting ECG segment is stored together with its valid length, allowing ECG and PPG to remain aligned in wall clock time. The final output is a compressed NPZ file for each participant and sensing location containing synchronized PPG and ECG windows, timestamps, and associated metadata.

\textbf{Ground truth generation and window filtering.} Ground truth heart rate is computed from the aligned ECG segment in each window. The ECG reference is recorded by the Polar chest strap at 130Hz, following the sensor output setting used during data collection. ECG is processed using NeuroKit2 \cite{makowski2021neurokit2}, with \texttt{ecg\_clean} and \texttt{ecg\_peaks} both using the \texttt{pantompkins1985} method \cite{pan1985real}. Heart rate is then derived from the median RR interval within the window. ECG quality is assessed using \texttt{ecg\_quality} with the \texttt{zhao2018} method and fuzzy approach \cite{zhao2018signal}. Only windows that pass the ECG-based quality checks are retained in the final windowed dataset. Specifically, windows are excluded if the ECG contains missing values, if the derived heart rate falls outside 40 to 180 beats per minute, if any RR interval falls outside 0.25 to 2.0s, if fewer than four beats are detected, if the coefficient of variation of RR intervals exceeds 0.2, or if the ECG quality label is marked as unacceptable. Invalid windows are removed entirely rather than retained with missing labels.

\textbf{Output format.} The final windowed dataset is stored in compressed NPZ format, with one file per participant and sensing location. Typical outputs include window start and end timestamps, synchronized PPG, synchronized ECG, ECG valid length, sampling rate, and ECG-derived heart rate labels.

\textbf{Compute.} The preparation pipeline was run on a personal MacBook Pro with an Apple M5 chip and 16~GB unified memory. Experiments were conducted on CPU only and did not use a GPU. A full run of the dataset preparation pipeline required approximately 5~hours to complete, and intermediate files occupied approximately 50~GB of storage.


\subsection{Heuristic Experiments}

We evaluate six heuristic baselines for PPG-based heart rate estimation, covering standard classes of traditional signal processing approaches. The six methods are the fast Fourier transform (FFT), HeartPy \cite{vangent2019heartpy}, multiscale peak and trough detection (MSPTD) \cite{bishop2018msptd}, peak width demodulation (PWD) \cite{li2010automatic}, qppgfast \cite{vest2018opensource}, and NeuroKit2 \cite{makowski2021neurokit2}. Together, they represent several commonly used families of PPG heart rate estimation algorithms, including frequency domain estimation, peak detection, and waveform based analysis. Because these methods do not rely on trainable parameters, they provide a useful reference for assessing the intrinsic signal quality of the dataset and the differences across sensing locations.

\textbf{Data pipeline.} All heuristic experiments are conducted on the 8-second windowed dataset. The evaluation pipeline supports both infrared (IR) and green PPG channels, and can be applied to either channel separately. The results reported in Table~\ref{tab:main_results_supplementary} use green PPG only. Each window contains 800 samples at 100~Hz. Before algorithm specific processing, each window is detrended using \texttt{scipy.signal.detrend} with \texttt{type="linear"}, and then filtered with a second-order Butterworth bandpass filter from 0.7 to 3.5~Hz using zero phase filtering implemented with \texttt{filtfilt}. No additional normalization or resampling is applied at this shared preprocessing stage.

\textbf{Evaluation protocol.} All heuristic experiments are conducted on the filtered 8-second windowed dataset described in Section~B.1. During evaluation, all retained windows are processed by each algorithm, and algorithm failures are recorded as NaN estimates. We report mean absolute error (MAE), root mean square error (RMSE), and Pearson correlation coefficient $\rho$. For each sensing location, metrics are first computed separately for each participant using all valid windows for that participant, and are then summarized across participants as mean $\pm$ standard deviation. Results are reported separately for the earring, ring, watch, and necklace.

\textbf{Implementation details.} Heuristic experiments are run through \texttt{runner.py}, with shared preprocessing implemented in \texttt{preprocess.py} and invoked automatically before evaluation. 
\begin{itemize}
    \item FFT is implemented by multiplying each 8-second segment by a Hann window, zero padding to 8192 points, computing the single sided real FFT spectrum, restricting peak search to 0.5 to 3.0~Hz, and applying parabolic interpolation around the peak bin. The resulting estimate is rejected if the inferred heart rate falls outside 30 to 180 beats per minute.
    
    \item  HeartPy is evaluated using the Python package implementation \cite{vangent2019heartpy}, with \texttt{scale\_data}, \texttt{enhance\_peaks}, and \texttt{process} using \texttt{bpmmin=40}, \texttt{bpmmax=180}, and \texttt{windowsize=0.75}. 

    \item NeuroKit2 is evaluated using \texttt{ppg\_clean} and \texttt{ppg\_peaks} with the Elgendi method \cite{elgendi2013systolic} and \texttt{correct\_artifacts=False}. 
    \item MSPTD is implemented in Python following the published method \cite{bishop2018msptd} and uses 6-second internal subwindows with 20\% overlap, peak refinement within a 5-sample neighborhood, mean interbeat interval aggregation, and a final heart rate range of 30 to 200 beats per minute. 
    \item PWD is implemented in Python following the delineator of Li et al. \cite{li2010automatic}, with a third-order Bessel low-pass filter at 25~Hz, additional moving average smoothing, median interbeat interval aggregation, and a final heart rate range of 30 to 200 beats per minute. 
    \item qppgfast is implemented in Python following the open source cardiovascular waveform toolbox of Vest et al. \cite{vest2018opensource}, using the standard slope sum function-based adaptive thresholding pipeline, rescaling to [-2000, 2000], median interbeat interval aggregation, and a final heart rate range of 30 to 200 beats per minute.
    
\end{itemize}
Unless otherwise noted, all methods are applied after the shared preprocessing described above. Some methods further include algorithm specific processing steps within their own implementations, as described above for HeartPy, NeuroKit2, PWD, and qppgfast.

\paragraph{Existing assets.}
We use WildPPG as an external reference dataset for the baseline comparison in Table~\ref{tab:main_results} and cite the original WildPPG publication. Most WildPPG results in Table~\ref{tab:main_results} are taken from the original publication; the only exception is NeuroKit, which we evaluate on WildPPG using our implementation and evaluation pipeline. WildPPG was accessed and used according to its stated access conditions and terms of use. WildPPG data are released under CC BY-NC-SA 4.0, and the WildPPG code is released under GPL-3.0. We do not redistribute or repackage WildPPG data or code as part of our release. Software libraries and baseline implementations used in our experiments are cited where applicable and used under their respective licenses.

\textbf{Compute.} Heuristic experiments were run on a MacBook Pro computer with an Apple M5 chip and 16~GB unified memory. Heuristic experiments were conducted on CPU only and did not use a GPU. A full run of the heuristic benchmark on the complete dataset required approximately 16~hours. 


\subsection{Supervised and Self-supervised Experiments}
\label{sec:supervised_ssl_experiments}

We evaluate five neural network models for PPG-based heart-rate estimation, covering convolutional, recurrent, and attention-based designs. All supervised models use only green-channel PPG downsampled to 25 Hz, a common sampling rate for wearable PPG modeling. The reported results from supervised models are all leave-one-subject-out based. The models include 1D ResNet and Fully Convolutional Network (FCN), two standard convolutional time-series baselines~\citep{wang2017time}; DeepConvLSTM (DCL), which combines convolutional feature extraction with recurrent temporal modeling~\citep{biswas2019cornet}; a CNN-LSTM variant with wider kernels and max-pooling; and a Transformer baseline that uses self-attention to model longer-range pulse dependencies. Together, these models compare major time-series modeling paradigms and their suitability for wearable PPG heart-rate estimation.

\paragraph{Data pipeline.}
Each 8-second window is resampled to 200 samples (25\,Hz) by polyphase resampling and z-score standardized per window. Windows with $\mathrm{hr}_\text{gt}\notin[30, 210]$\,bpm or any non-finite sample are discarded. An optional 0.7--3.5\,Hz Butterworth filter is available in the code to detrend the input PPG signal, but we did not use it in the reported results. For multi-site fusion, the four per-device window streams are aligned by nearest start timestamp with a 500\,ms tolerance and cached once per participant.

\paragraph{Splits and protocol.}
We use leave-one-subject-out (LOSO) over 20 participants: each fold holds out one participant for testing, and the remaining 19 are pooled and split 80/20 at the window level into train and validation. We report MAE (bpm), RMSE (bpm), and Pearson $\rho$ (mean $\pm$ std) across participants for all experiments.

\paragraph{Backbones.}
All models take input of shape $(B, 200, C)$ with $C{\in}\{1,4,8\}$ for single-position, green-only multi-site, and green$+$IR/ACC multi-site, respectively, and emit a single bpm prediction trained with L1 loss. The backbone implementations are adapted from the public WildPPG codebase~\cite{meier2024wildppg}. 

\begin{table}[h]
    \centering
    \small
    \caption{Backbone architectures used as supervised baselines. All variants share input shape, output head, and L1 loss.}
    \label{tab:backbones}
    \setlength{\tabcolsep}{4pt}
    \renewcommand{\arraystretch}{1.0}
    \tiny
    \begin{tabular}{@{}llll@{}}
        \toprule
        \textbf{Backbone} & \textbf{Core blocks} & \textbf{Width} & \textbf{Notable settings} \\
        \midrule
        FCN & 3 Conv1d--BN--ReLU--MaxPool & 32 / 64 / 128 & kernel 8, dropout 0.35 \\
        DCL & 4 Conv2d ($5{\times}1$) $\to$ 2-layer LSTM & 64 conv, 128 LSTM & dropout 0.5 between stages \\
        CNN--LSTM & 2 Conv1d (kernel 32, 16) $\to$ 2-layer LSTM & 32 / 64 conv, 128 LSTM & dropout 0.2 \\
        Transformer & 4 encoder blocks & $d{=}128$, 4 heads, MLP 64 & learnable \texttt{[CLS]}, dropout 0.1 \\
        ResNet1D & Stem conv $+$ 8 residual blocks & 32 base filters & kernel 5, stride 2 \\
        \bottomrule
    \end{tabular}
\end{table}

\paragraph{Optimization and hyperparameter selection.}
We train with Adam at $5\times10^{-4}$, batch size 128, for up to 20 epochs, with early stopping after 5 epochs without validation-MAE improvement. Optimization hyperparameters were chosen by a single pilot sweep on one LOSO fold (learning rate over $\{1,3,5\}\times10^{-4}, \{1,3\}\times10^{-3}$; batch size over $\{64, 128, 256\}$; patience over $\{3, 5, 10\}$) and then frozen across all backbones, folds, and seeds; no per-fold retuning was performed. Architecture hyperparameters are fixed to each method's reference values.

\paragraph{Self-supervised pretraining.}
SSL baselines (SimCLR, BYOL) use the CNN--LSTM as encoder with a two-layer projection head of width 128. Two views are generated per window via additive Gaussian noise ($\sigma{=}0.05$) and per-sample amplitude scaling ($s\sim\mathcal{U}[0.8, 1.2]$). Pretraining runs for 60 epochs with Adam at $3\times10^{-3}$, batch size 256, cosine decay; SimCLR uses InfoNCE with $\tau{=}0.07$, BYOL uses an EMA target with momentum 0.99. The encoder is then frozen, and a linear regression head is fit for 60 epochs at $1\times10^{-3}$ under the same LOSO splits. The held-out test participant is excluded from both stages.

\paragraph{Hardware and compute.}
Experiments were run on a GPU cluster with NVIDIA H100 GPUs. For the heaviest supervised backbone, the Transformer, a full leave-one-subject-out run on one device dataset required approximately 6 GPU-hours. The complete supervised benchmark required approximately 60 GPU-hours for one random seed, and the complete self-supervised benchmark required an additional 100 GPU-hours.


\subsection{Multi-site Experiment}
\label{multisite}
\paragraph{Data pipeline.}
Multi-site fusion takes the four per-device window streams (earring, ring, watch, necklace) and stacks the green PPG channels into a single $C{=}4$ input tensor per window. Because each device begins and ends recording at slightly different timestamps, we align windows across devices before stacking: the device with the fewest windows acts as the reference, and for every reference window with start timestamp $t$ we select the window from each remaining device whose $t_0$ minimises $|t_0 - t|$, provided the residual is at most 500\,ms. Reference windows for which any device fails this tolerance are dropped. The aligned tensor is cached once per participant as \texttt{aligned\_4device.npz} and reused across every multi-site run, so that all backbones see the exact same set of fused windows.
 
\paragraph{Setup.}
Backbones, optimization, splits, and reporting follow Section~\ref{sec:supervised_ssl_experiments} unchanged; the only difference is the input channel count, $C{=}4$ instead of $C{=}1$. We compare to single-position baselines by re-running each backbone independently per position under the same LOSO splits, so that the reported gain (or loss) from fusion is attributable to the additional channels rather than to differences in seeds, splits, or training schedule.

\paragraph{Hardware and compute.}
Experiments were run on a GPU cluster with NVIDIA H100 GPUs. The complete multi-site experiment took approximately 60 GPU-hours.

\subsection{Multimodal Experiment}
\paragraph{Setup.}
We run two separate multimodal ablations on top of the multi-site setup, each adding a different second per-device modality and retraining each backbone from scratch at $C{=}8$. Both share the multi-site data pipeline of Section~\ref{multisite}, including the 500\,ms cross-device alignment, and both follow the optimizer, splits, and early-stopping criterion of Section~\ref{sec:supervised_ssl_experiments}.
 
\paragraph{PPG green $+$ IR.}
The first ablation adds a second optical wavelength: the infrared PPG channel is stacked alongside the green PPG channel at each of the four positions, yielding $C{=}8$. No additional preprocessing is applied to the IR channel beyond the per-window resampling and z-score standardization used for green. The aligned tensor is cached as \texttt{aligned\_8channel.npz}. Comparing this run against the green-only multi-site run isolates the contribution of the additional wavelength.
 
\paragraph{PPG green $+$ accelerometer.}
The second ablation adds device motion: the vertical-axis accelerometer channel \texttt{accel\_z} is stacked alongside green PPG at each of the four positions, again yielding $C{=}8$. The accelerometer signal is resampled and z-score standardized per window in exactly the same way as the PPG channels, so that the model receives all eight channels on the same numerical scale. The aligned tensor is cached as \texttt{aligned\_8channel\_accel.npz}. Comparing this run against the green-only multi-site run isolates the contribution of motion as a side-information modality, and comparing it against the green$+$IR run directly contrasts the value of an additional wavelength against the value of an additional sensor.

\paragraph{Hardware and compute.}
Experiments were run on a GPU cluster with NVIDIA H100 GPUs. The complete multimodal experiment took approximately 100 GPU-hours.

\section{Supplementary Results}

\subsection{Main Baseline Results with Standard Deviation}
Table~\ref{tab:main_results_supplementary} provides detailed results with standard deviation for Table~\ref{tab:main_results} that we did not have space to include in the main paper.

\begin{table*}[h]
\centering
\caption{Baseline results on our Multi-site PPG Dataset with std, as a complementary table to Table~\ref{tab:main_results}. Lower is better for MAE/RMSE, higher is better for $\rho$ (range -1 to 1).}
\label{tab:main_results_supplementary}
\renewcommand{\arraystretch}{1.15}
\resizebox{\textwidth}{!}{%
\begin{tabular}{lccc|ccc|ccc|ccc}
\toprule
\multirow{3}{*}{Method}
& \multicolumn{12}{c}{\textbf{Multi-site PPG Dataset Results (with std)}} \\
\cmidrule(lr){2-13}
& \multicolumn{3}{c|}{Earring}
& \multicolumn{3}{c|}{Ring}
& \multicolumn{3}{c|}{Watch}
& \multicolumn{3}{c}{Necklace} \\
\cmidrule(lr){2-4} 
\cmidrule(lr){5-7} 
\cmidrule(lr){8-10} 
\cmidrule(lr){11-13}
& MAE$\downarrow$ & RMSE$\downarrow$ & $\rho\uparrow$
& MAE$\downarrow$ & RMSE$\downarrow$ & $\rho\uparrow$
& MAE$\downarrow$ & RMSE$\downarrow$ & $\rho\uparrow$
& MAE$\downarrow$ & RMSE$\downarrow$ & $\rho\uparrow$ \\
\midrule

\multicolumn{13}{l}{\textit{\textbf{Heuristic}}} \\

NeuroKit
& $\mathbf{2.30 \pm 1.28}$ & $\mathbf{5.28 \pm 4.15}$ & $\mathbf{0.933 \pm 0.164}$
& $7.54 \pm 2.14$ & $13.64 \pm 3.50$ & $0.471 \pm 0.221$
& $12.43 \pm 1.92$ & $18.84 \pm 2.69$ & $0.221 \pm 0.159$
& $14.58 \pm 2.75$ & $20.72 \pm 3.36$ & $0.178 \pm 0.156$ \\

HeartPy
& $2.60 \pm 1.48$ & $6.43 \pm 4.50$ & $0.903 \pm 0.184$
& $9.44 \pm 3.54$ & $16.91 \pm 5.06$ & $0.459 \pm 0.190$
& $15.68 \pm 3.77$ & $23.74 \pm 3.87$ & $0.222 \pm 0.151$
& $20.05 \pm 5.05$ & $28.67 \pm 6.10$ & $0.145 \pm 0.122$ \\

qppgfast
& $3.36 \pm 2.69$ & $7.67 \pm 5.68$ & $0.871 \pm 0.248$
& $11.94 \pm 4.27$ & $19.74 \pm 5.54$ & $0.255 \pm 0.204$
& $15.09 \pm 2.36$ & $22.02 \pm 2.62$ & $0.194 \pm 0.141$
& $17.84 \pm 4.11$ & $25.16 \pm 4.72$ & $0.121 \pm 0.096$ \\

MSPTD
& $2.47 \pm 1.37$ & $6.13 \pm 4.29$ & $0.906 \pm 0.165$
& $10.09 \pm 2.48$ & $17.13 \pm 3.70$ & $0.331 \pm 0.204$
& $16.54 \pm 3.38$ & $23.58 \pm 3.54$ & $0.125 \pm 0.128$
& $19.72 \pm 3.43$ & $26.33 \pm 3.75$ & $0.109 \pm 0.139$ \\

PWD
& $2.71 \pm 1.55$ & $6.71 \pm 4.78$ & $0.886 \pm 0.172$
& $11.68 \pm 2.96$ & $19.81 \pm 4.16$ & $0.194 \pm 0.184$
& $17.59 \pm 3.47$ & $25.16 \pm 3.84$ & $0.069 \pm 0.114$
& $19.13 \pm 3.91$ & $25.62 \pm 4.35$ & $0.095 \pm 0.125$ \\

FFT
& $2.67 \pm 1.29$ & $6.98 \pm 4.22$ & $0.875 \pm 0.159$
& $11.93 \pm 2.97$ & $20.55 \pm 4.06$ & $0.238 \pm 0.183$
& $19.19 \pm 4.12$ & $27.00 \pm 3.98$ & $0.093 \pm 0.116$
& $22.46 \pm 3.87$ & $30.01 \pm 3.74$ & $0.098 \pm 0.128$ \\

\midrule

\multicolumn{13}{l}{\textit{\textbf{Supervised}}} \\

1D ResNet
& $2.44 \pm 1.28$ & $5.35 \pm 4.42$ & $0.879 \pm 0.181$
& $5.47 \pm 2.43$ & $\mathbf{10.10 \pm 4.37}$ & $\mathbf{0.679 \pm 0.209}$
& $8.74 \pm 2.30$ & $13.94 \pm 3.97$ & $0.439 \pm 0.192$
& $9.01 \pm 2.91$ & $13.26 \pm 3.85$ & $\mathbf{0.524 \pm 0.120}$ \\

DCL
& $2.34 \pm 1.38$ & $5.63 \pm 4.86$ & $0.867 \pm 0.197$
& $\mathbf{5.13 \pm 2.27}$ & $10.13 \pm 4.20$ & $0.676 \pm 0.203$
& $8.50 \pm 2.21$ & $13.73 \pm 3.94$ & $0.425 \pm 0.197$
& $\mathbf{8.68 \pm 3.02}$ & $\mathbf{13.07 \pm 3.99}$ & $0.521 \pm 0.118$ \\

FCN
& $2.66 \pm 1.38$ & $5.45 \pm 4.17$ & $0.882 \pm 0.166$
& $6.12 \pm 2.65$ & $10.29 \pm 4.18$ & $0.650 \pm 0.196$
& $9.39 \pm 3.03$ & $13.94 \pm 4.40$ & $0.423 \pm 0.176$
& $9.35 \pm 3.35$ & $13.09 \pm 4.11$ & $0.497 \pm 0.134$ \\

CNN-LSTM
& $2.43 \pm 1.47$ & $5.96 \pm 5.13$ & $0.858 \pm 0.198$
& $5.33 \pm 2.49$ & $10.37 \pm 4.47$ & $0.661 \pm 0.203$
& $\mathbf{8.37 \pm 2.33}$ & $\mathbf{13.55 \pm 3.93}$ & $\mathbf{0.458 \pm 0.179}$
& $8.74 \pm 3.17$ & $13.21 \pm 4.14$ & $0.497 \pm 0.139$ \\

Transformer
& $2.79 \pm 1.60$ & $6.47 \pm 5.24$ & $0.841 \pm 0.217$
& $6.71 \pm 2.69$ & $11.90 \pm 4.68$ & $0.539 \pm 0.219$
& $9.74 \pm 2.67$ & $14.74 \pm 4.17$ & $0.332 \pm 0.171$
& $10.83 \pm 3.69$ & $14.49 \pm 4.29$ & $0.349 \pm 0.115$ \\

\midrule
\multicolumn{13}{l}{\textit{\textbf{Self-supervised}}} \\

SimCLR
& $3.76 \pm 1.59$ & $7.32 \pm 4.21$ & $0.823 \pm 0.144$
& $9.10 \pm 3.06$ & $13.70 \pm 4.41$ & $0.332 \pm 0.163$
& $10.72 \pm 3.80$ & $14.49 \pm 4.99$ & $0.203 \pm 0.109$
& $11.25 \pm 3.82$ & $14.88 \pm 4.57$ & $0.236 \pm 0.108$ \\

BYOL
& $3.46 \pm 1.57$ & $7.26 \pm 4.60$ & $0.822 \pm 0.163$
& $7.83 \pm 3.16$ & $12.44 \pm 4.56$ & $0.475 \pm 0.200$
& $11.63 \pm 3.90$ & $15.31 \pm 4.79$ & $0.234 \pm 0.160$
& $11.13 \pm 3.70$ & $14.53 \pm 4.37$ & $0.317 \pm 0.124$ \\

\bottomrule
\end{tabular}%
}
\end{table*}

\section{Additional Experiment Results}

\subsection{Results Across Participants}
\label{results_across_skintone}

\begin{figure}[ht]
    \centering
    \includegraphics[width=1\linewidth]{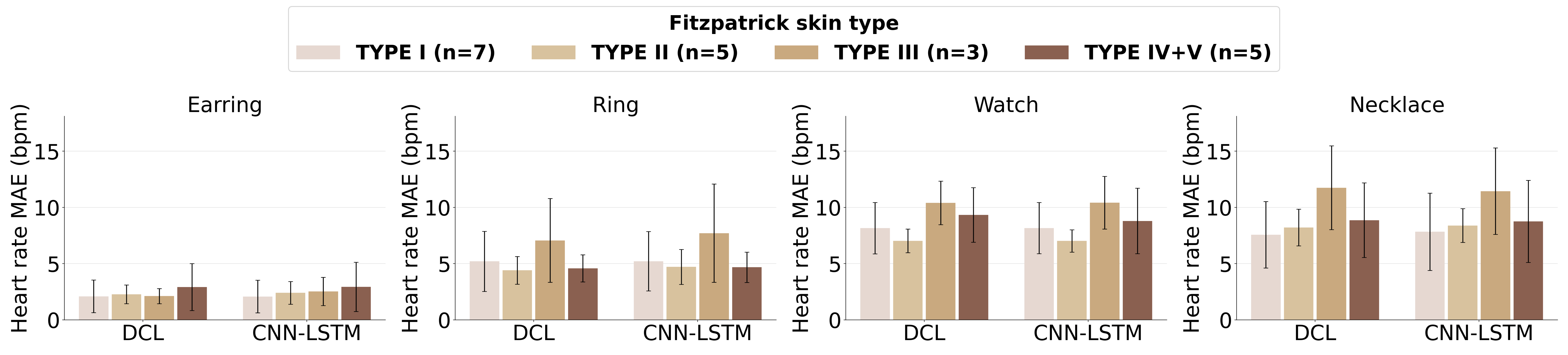}
    \caption{\small{Skin color type–wise comparison of heart-rate MAE across models for each device.}}
    \label{fig:skin_color_baseline_results}
    \vskip -1em
\end{figure}

We compute the average MAE across participants with different Fitzpatrick skin types. Figure~\ref{fig:skin_color_baseline_results} shows heart-rate estimation accuracy across the four wearable locations for participants with different skin tones. We do not observe a clear relationship between skin tone and estimation error in our dataset. Although reduced PPG accuracy for darker skin tones has been reported in prior work~\citep{koerber2023accuracy}, errors in our in-the-wild recordings appear to be dominated by motion artifacts, as discussed in Section~\ref{motion_effect}. In daily-life settings, participants’ activities and behaviors can introduce substantially different levels of motion, which may have a stronger effect on PPG performance than skin tone in this dataset.

\subsection{Multi-site PPG-Motion Fusion}
\label{multisite_ppg_motion_fusion}

We discussed in Section~\ref{ppg_motion_fusion} how adding accelerometer data to green-channel PPG can improve heart-rate estimation for individual wearable devices. In single-device models, accelerometer input generally improves performance for the ring and watch, with the most noticeable gain on motion-sensitive sites such as the watch, where MAE decreases by about 0.3 bpm. This suggests that accelerometer signals provide useful motion context that helps models compensate for daily-life motion artifacts.

We further evaluate multi-site PPG--motion fusion across three backbones, with results shown in Tables~\ref{tab:accl_fusion_resnet}--\ref{tab:accl_fusion_cnn_lstm}. We consider eight device configurations: four single-device settings (earring, ring, watch, necklace), three adjacent two-device fusions (earring+ring, ring+watch, watch+necklace), and full four-device fusion. For each configuration, we compare two modality settings: \textbf{PPG only}, using green-channel PPG from the selected devices, and \textbf{PPG + Accelerometer}, using green-channel PPG together with accelerometer $z$-axis data. The evaluated backbones are 1D ResNet, DCL, and CNN--LSTM, each trained from scratch using the same hyperparameters described in Table~\ref{tab:backbones}.

\begin{table}[h]
\centering
\caption{1D ResNet cross-modality results across 8 device configurations. Values are mean $\pm$ std across 20 LOSO folds.}
\label{tab:accl_fusion_resnet}
\setlength{\tabcolsep}{2pt}
\renewcommand{\arraystretch}{0.95}
\tiny
\begin{tabular}{l|ccc|ccc}
\toprule
\textbf{1D ResNet}
 & \multicolumn{3}{c|}{PPG only} & \multicolumn{3}{c}{PPG + Accelerometer ($z$ axis)} \\
 & MAE$\downarrow$ & RMSE$\downarrow$ & $\rho\uparrow$ & MAE$\downarrow$ & RMSE$\downarrow$ & $\rho\uparrow$ \\
\midrule
\multicolumn{7}{l}{\textit{Single device}} \\
Earring            & $1.84\pm0.71$ & $3.19\pm1.61$ & $0.94\pm0.08$ & $1.93\pm0.74$ & $3.26\pm1.71$ & $0.94\pm0.09$ \\
Ring               & $4.55\pm2.04$ & $7.85\pm3.02$ & $0.74\pm0.14$ & $4.42\pm1.71$ & $7.67\pm2.67$ & $0.75\pm0.13$ \\
Watch              & $7.03\pm2.43$ & $10.59\pm3.25$ & $0.50\pm0.19$ & $6.74\pm2.27$ & $10.00\pm3.08$ & $0.57\pm0.18$ \\
Necklace           & $7.73\pm2.72$ & $11.17\pm3.33$ & $0.48\pm0.13$ & $7.79\pm3.22$ & $11.11\pm3.96$ & $0.51\pm0.17$ \\
\midrule
\multicolumn{7}{l}{\textit{2-device fusion}} \\
Earring + Ring     & $1.82\pm0.65$ & $3.18\pm1.57$ & $0.95\pm0.07$ & $1.92\pm0.68$ & $3.25\pm1.64$ & $0.94\pm0.08$ \\
Ring + Watch       & $4.48\pm1.97$ & $7.82\pm2.89$ & $0.73\pm0.15$ & $4.38\pm1.72$ & $7.56\pm2.71$ & $0.75\pm0.13$ \\
Watch + Necklace   & $6.90\pm2.32$ & $10.36\pm3.19$ & $0.53\pm0.18$ & $6.79\pm2.27$ & $10.08\pm3.14$ & $0.57\pm0.17$ \\
\midrule
\multicolumn{7}{l}{\textit{4-device fusion}} \\
4-device           & $1.89\pm0.69$ & $3.22\pm1.62$ & $0.95\pm0.08$ & $2.00\pm0.73$ & $3.44\pm1.77$ & $0.94\pm0.09$ \\
\bottomrule
\end{tabular}
\end{table}

\begin{table}[h]
\centering
\caption{DeepConvLSTM (DCL) cross-modality results across 8 device configurations. Values are mean $\pm$ std across 20 LOSO folds.}
\label{tab:accl_fusion_DCL}
\setlength{\tabcolsep}{2pt}
\renewcommand{\arraystretch}{0.95}
\tiny
\begin{tabular}{l|ccc|ccc}
\toprule
\textbf{DCL}
 & \multicolumn{3}{c|}{PPG only} & \multicolumn{3}{c}{PPG + Accelerometer ($z$ axis)} \\
 & MAE$\downarrow$ & RMSE$\downarrow$ & $\rho\uparrow$ & MAE$\downarrow$ & RMSE$\downarrow$ & $\rho\uparrow$ \\
\midrule
\multicolumn{7}{l}{\textit{Single device}} \\
Earring            & $1.65\pm0.67$ & $3.04\pm1.69$ & $0.95\pm0.07$ & $1.61\pm0.63$ & $2.90\pm1.55$ & $0.95\pm0.07$ \\
Ring               & $4.23\pm2.17$ & $7.67\pm3.15$ & $0.74\pm0.15$ & $4.18\pm1.73$ & $7.74\pm2.70$ & $0.75\pm0.14$ \\
Watch              & $6.90\pm2.37$ & $10.58\pm3.18$ & $0.50\pm0.19$ & $6.58\pm2.09$ & $10.11\pm2.82$ & $0.57\pm0.17$ \\
Necklace           & $7.61\pm2.86$ & $11.14\pm3.53$ & $0.48\pm0.16$ & $7.91\pm3.34$ & $11.26\pm3.95$ & $0.52\pm0.15$ \\
\midrule
\multicolumn{7}{l}{\textit{2-device fusion}} \\
Earring + Ring     & $1.65\pm0.79$ & $2.86\pm1.77$ & $0.95\pm0.09$ & $1.60\pm0.66$ & $2.79\pm1.58$ & $0.95\pm0.07$ \\
Ring + Watch       & $4.42\pm2.26$ & $8.06\pm3.24$ & $0.73\pm0.15$ & $4.20\pm1.65$ & $7.72\pm2.50$ & $0.75\pm0.14$ \\
Watch + Necklace   & $6.13\pm2.33$ & $9.78\pm3.21$ & $0.59\pm0.20$ & $6.78\pm2.51$ & $10.34\pm3.30$ & $0.58\pm0.18$ \\
\midrule
\multicolumn{7}{l}{\textit{4-device fusion}} \\
4-device           & $1.65\pm0.72$ & $2.94\pm1.71$ & $0.95\pm0.09$ & $1.62\pm0.65$ & $2.82\pm1.58$ & $0.95\pm0.07$ \\
\bottomrule
\end{tabular}
\end{table}

\begin{table}[h]
\centering
\caption{CNN-LSTM cross-modality results across 8 device configurations. Values are mean $\pm$ std across 20 LOSO folds.}
\label{tab:accl_fusion_cnn_lstm}
\setlength{\tabcolsep}{2pt}
\renewcommand{\arraystretch}{0.95}
\tiny
\begin{tabular}{l|ccc|ccc}
\toprule
 \textbf{CNN-LSTM}
 & \multicolumn{3}{c|}{PPG only} & \multicolumn{3}{c}{PPG + Accelerometer ($z$ axis)} \\
 & MAE$\downarrow$ & RMSE$\downarrow$ & $\rho\uparrow$ & MAE$\downarrow$ & RMSE$\downarrow$ & $\rho\uparrow$ \\
\midrule
\multicolumn{7}{l}{\textit{Single device}} \\
Earring            & $1.67\pm0.67$ & $3.18\pm1.72$ & $0.95\pm0.07$ & $1.68\pm0.66$ & $3.13\pm1.70$ & $0.95\pm0.07$ \\
Ring               & $4.42\pm2.17$ & $8.03\pm3.25$ & $0.72\pm0.15$ & $4.42\pm2.07$ & $8.02\pm3.05$ & $0.73\pm0.14$ \\
Watch              & $6.96\pm2.36$ & $10.67\pm3.18$ & $0.49\pm0.19$ & $6.74\pm2.23$ & $10.23\pm3.07$ & $0.55\pm0.18$ \\
Necklace           & $7.59\pm2.85$ & $11.26\pm3.49$ & $0.47\pm0.13$ & $7.63\pm3.14$ & $11.05\pm3.83$ & $0.51\pm0.15$ \\
\midrule
\multicolumn{7}{l}{\textit{2-device fusion}} \\
Earring + Ring     & $1.64\pm0.65$ & $3.03\pm1.68$ & $0.95\pm0.07$ & $1.68\pm0.64$ & $3.14\pm1.69$ & $0.95\pm0.07$ \\
Ring + Watch       & $4.56\pm2.10$ & $8.30\pm3.08$ & $0.70\pm0.16$ & $4.50\pm1.93$ & $8.19\pm2.86$ & $0.72\pm0.13$ \\
Watch + Necklace   & $6.26\pm2.24$ & $9.94\pm3.18$ & $0.58\pm0.15$ & $6.41\pm2.44$ & $9.89\pm3.40$ & $0.60\pm0.17$ \\
\midrule
\multicolumn{7}{l}{\textit{4-device fusion}} \\
4-device           & $1.66\pm0.63$ & $3.08\pm1.60$ & $0.95\pm0.07$ & $1.70\pm0.68$ & $3.21\pm1.72$ & $0.94\pm0.08$ \\
\bottomrule
\end{tabular}
\end{table}

For multi-device fusion, the benefits of accelerometer input are more mixed. Adding accelerometer data improves ring + watch and watch + necklace fusion, but slightly degrades earring + ring and four-device fusion. This suggests that motion information is most helpful when the PPG channels are strongly affected by motion, but may add unnecessary complexity when the selected PPG sites already provide strong signals. Overall, these results show that our dataset supports studying both multi-site PPG fusion and cross-modality fusion, and that accelerometer signals provide meaningful complementary information for motion-robust heart-rate estimation.

\subsection{Multi-wavelength PPG Fusion}
\label{multi_wave_ppg_fusion}

\begin{table}[h]
\caption{Multi-wavelength fusion results using 1D ResNet. Values are mean$\pm$std across 20 leave-one-subject-out folds.}
\label{tab:multi_wavelength_fusion}
\centering
\setlength{\tabcolsep}{2pt}
\renewcommand{\arraystretch}{0.95}
\small
\begin{tabular}{l|ccc|ccc}
\toprule
                 & \multicolumn{3}{c|}{PPG green only} & \multicolumn{3}{c}{PPG green + IR} \\
                 & MAE$\downarrow$ & RMSE$\downarrow$ & $\rho\uparrow$ & MAE$\downarrow$ & RMSE$\downarrow$ & $\rho\uparrow$ \\
\midrule
\multicolumn{7}{l}{\textit{Single device}} \\
Earring   & 1.84$\pm$0.71 & 3.19$\pm$1.61  & 0.944$\pm$0.085 & 1.81$\pm$0.65 & 3.12$\pm$1.59  & 0.948$\pm$0.073 \\
Ring      & 4.55$\pm$2.04 & 7.85$\pm$3.02  & 0.735$\pm$0.144 & 4.46$\pm$2.01 & 7.77$\pm$3.09  & 0.738$\pm$0.159 \\
Watch     & 7.03$\pm$2.43 & 10.59$\pm$3.25 & 0.498$\pm$0.188 & 6.84$\pm$2.31 & 10.39$\pm$3.15 & 0.508$\pm$0.199 \\
Necklace  & 7.73$\pm$2.72 & 11.17$\pm$3.33 & 0.480$\pm$0.129 & 7.77$\pm$2.70 & 11.17$\pm$3.35 & 0.487$\pm$0.114 \\
\midrule
\multicolumn{7}{l}{\textit{4-device fusion}} \\
4-device  & 1.94$\pm$0.73 & 3.31$\pm$1.73  & 0.942$\pm$0.088 & 1.98$\pm$0.74 & 3.44$\pm$1.79  & 0.939$\pm$0.087 \\
\bottomrule
\end{tabular}
\end{table}

The multi-wavelength fusion results show that adding IR PPG to green PPG provides small but consistent improvements for most single-device settings. For the earring, ring, and watch, green+IR fusion reduces MAE and RMSE and slightly improves Pearson correlation, with the watch showing the largest MAE reduction from 7.03 bpm to 6.84 bpm. This suggests that IR can provide complementary information to green PPG, especially for more motion-sensitive or lower-quality sensing sites. However, the necklace does not benefit from IR fusion, with MAE slightly increasing from 7.73 bpm to 7.77 bpm, likely because its pendant form factor introduces unstable skin contact that limits the usefulness of additional optical channels. In the four-device setting, adding IR also slightly worsens performance, suggesting that simply increasing the number of input channels does not always improve accuracy and may introduce additional noise or modeling complexity. Overall, these results indicate that multi-wavelength PPG can be beneficial for individual wearable sites, but its effectiveness depends on the sensing location and signal quality.


\end{document}